\documentclass[twocolumn,showpacs,preprintnumbers,amsmath,amssymb,nofootinbib,APS]{revtex4}

\usepackage{graphicx}
\usepackage{dcolumn}
\usepackage{bm}

\newcommand{\be}{\begin{equation}}
\newcommand{\ee}{\end{equation}}

\begin{document}

\title{ {\bf Black hole radiance, short distances, and TeV gravity}}
\author{Iv\'{a}n Agull\'{o}, Jos\'{e} Navarro-Salas}\email{ivan.agullo@uv.es;
jnavarro@ific.uv.es} \affiliation{ {\footnotesize Departamento de
F\'{\i}sica Te\'orica and
    IFIC, Centro Mixto Universidad de
    Valencia-CSIC \\
 Universidad de Valencia, Burjassot-46100, Valencia, Spain}}
\author{Gonzalo J. Olmo}\email{olmoalba@uwm.edu}
\affiliation{
    {\footnotesize  Physics Department, University of Wisconsin-Milwaukee,
P.O. Box 413, Milwaukee, Wisconsin 53201 USA }}

\date{April 6, 2006}


\begin{abstract}

 Using a derivation of black hole radiance in terms of two-point functions one can provide
a quantitative estimate  of the contribution of short distances to
the spectrum. Thermality is preserved for black holes with $\kappa
l_P <<1$. However, deviations from the Planckian spectrum can be
found for mini black holes in TeV gravity scenarios, even before
reaching the Planck phase.

\end{abstract}

\maketitle

Black hole radiance  is one of the most important consequences of
combining general relativity and quantum mechanics. Using quantum
field theory in curved spacetime Hawking \cite{hawk1} showed that
a black hole emits thermal radiation. The derivation involves
considering arbitrarily high frequency wave-packets in the
intermediate states of the derivation. Any out-going Hawking
quanta with finite energy at infinity will have an  exponentially
increasing frequency when it is propagated backwards in time and
measured by a free-falling observer at the horizon. The crucial
role played by these ultrahigh frequencies in the derivation of
the Planckian spectrum, or equivalently, the short-distance
behavior of the free field considered,  was stressed in
\cite{hooft, jacobson9193}. This question has been mainly analyzed
using sonic black hole models with modified high frequency
dispersion relations \cite{unruh95, bmps-cj} so as to eliminate
ultrashort wavelength modes. In doing so one must assume the
existence of a preferred frame.  Such a frame is naturally
identified with the rest frame of the atoms of the fluid and the
modified dispersion relations come from effects of its microscopic
structure. The result is that, even with a drastic change of the
theory, thermality is essentially unaffected if the black hole
scale is far from the underlying microscopic scale. This does not
exclude that, for small black holes, with size not too far from
the fundamental length scale, the standard Planckian spectrum can
be modified.

The purpose of this paper is to analyze this issue, in a purely
gravitational context, in terms of two-point functions instead of
dispersion relations. This way the short-distance contribution to
the spectrum  can be evaluated in a more explicit  way. We focus
our analysis on the situation where non-trivial deviations from
thermality can be found, even before reaching the late stages
(Planck scale) of the evaporation. Therefore we shall pay
particular attention to mini black holes considered recently
\cite{giddings-thomas02, dimopoulos-landsberg02, bhaccelerators}
in TeV gravity scenarios. The existence of extra dimensions gives
hope to the possibility that the fundamental Planck mass could be
TeV order \cite{extradimensions}. This, in turn, opens the
viability of producing black holes by high energy collisions
\cite{banks-fischler99} (as in the LHC or in cosmic ray
scattering) and detecting the Standard Model quanta of Hawking
radiation \cite{emparan-horowitz-myers}. Such black holes need to
be very small (less than the typical length of extra dimensions)
and above
 the fundamental Planck scale to apply semiclassical gravity. In
this scenario measurable deviations from thermality can arise due
to unknown physics at ultrashort distances.

 The mean particle number produced in the gravitational collapse
of a rotating black hole is \be \label{planck} \langle N_i \rangle
= \frac{\Gamma_i}{e^{2\pi{\kappa}^{-1}(w_i-m\Omega_{H})}-
{(-)}^{2s}} \ , \ee where $\kappa$ and $\Omega_{H}$ are the
surface gravity and the angular velocity, respectively, of the
black hole horizon. The $\Gamma_i$ are grey-body factors,
associated to a wave-packet $i$-mode (sharply peaked around the
frequency $\omega_i$) of a given particle species of spin $s$,
 and $m$ is the
axial angular momentum of the emitted particle. Up to grey-body
coefficients the spectrum is purely Planckian  with the chemical
potential term $m\Omega_{H}$. Note that the scale of
(\ref{planck}) is essentially given by the (classical)  surface
gravity $\kappa$ of the black hole. Moreover the radiation is
exactly thermal in the sense that there is no correlation between
different modes ($i\neq j$) \be \label{uncorrelated}\langle N_i
N_j \rangle = \langle N_i\rangle \langle N_j\rangle \ . \ee When
the modes coincide ($i=j$) the result is consistent with the
thermal probability distribution and the state of radiation  is
indeed described by a thermal density matrix \cite{parker75,
wald75} (see also \cite{waldbook, icp05}).

The above results are consequence of the evaluation of the
late-time Bogolubov coefficients in a gravitational collapse. The
expansion of a field in two different sets of positive frequency
modes: $u^{in}_j(x)$ (in the past infinity) and $u^{out}_j(x)$ (in
the future infinity) leads to a relation for the corresponding
creation and annihilation operators:  $ a_i^{out}=\sum_j (\alpha^*
_{ij}a_j^{in}-\beta ^*_{ij}a_j^{in \dagger})$. When the
coefficients $\beta _{ij}$ do not vanish  the vacuum states
$|in\rangle$ and $|out\rangle$ do not coincide and, therefore, the
number of particles measured in the $i^{th}$ mode by an ``out''
observer,  in the state $|in\rangle$ is given by
 $\langle in| N_i^{out}|in \rangle = \sum_k
|\beta_{ik}|^2$. Moreover the correlations for $i\neq j$ are given
by $\langle in| N_i N_j|in \rangle = (\sum_k |\beta_{ik}|^2)
(\sum_k |\beta_{jk}|^2)+ |\sum_k\beta_{ik}\beta_{jk}^*|^2 +
|\sum_k\alpha_{ik}\beta_{jk}|^2$. The use of the above relations
and the  explicit evaluation of the matrices $\beta_{ij}$ and
$\alpha_{ij}$ at late-times, which always involves to consider
intermediate ultrahigh frequency modes (due to the exponential
redshift associated to the black hole horizon), leads to the
thermal results (\ref{planck}) and (\ref{uncorrelated}).

Within the standard analysis in terms of Bogolubov coefficients it
is not easy to evaluate explicitly how ultrahigh frequencies or,
equivalently, ultrashort distances  contribute to generate the
thermal spectrum. However, it is not difficult to rederive the
Hawking effect in such a way that the contribution of
short-distance physics  can be explicitly worked out. Let us
assume, for the sake of simplicity, that $\phi$ is a massless,
neutral and minimally coupled scalar field.
One
can easily verify that the number operator can be obtained from
the following  projection
\begin{eqnarray}\label{eq:ai+aj}
{a^{out}}^\dagger_i a^{out}_j&=& \int_\Sigma d\Sigma_1 ^\mu
d\Sigma_2 ^\nu
[u^{out}_i(x_1){\buildrel\leftrightarrow\over{\partial}}_\mu
][u^{out*}_j(x_2){\buildrel\leftrightarrow\over{\partial}}_\nu ] \nonumber \\
&&\times (\phi (x_1)\phi (x_2)- \langle out|\phi (x_1)\phi
(x_2)|out\rangle)
 \ , \ \ \ \end{eqnarray}
where $\Sigma $ represents a suitable initial value hypersurface
and the two-point expectation value has the form $\langle out|\phi
(x_1)\phi (x_2)|out\rangle=\hbar\sum_k
u_k^{out}(x_1){u_k^{out}}^*(x_2)$. Therefore, the number of
particles in the $i^{th}$ mode measured by the ``out'' observer in
the ``in'' vacuum is given by $\langle in|N_i|in\rangle \equiv
\langle in|N_{ii}|in\rangle$, where $N_{ij}\equiv\hbar^{-1}
{a^{out}}^\dagger_i a^{out}_j $, and it can be evaluated using the
above expression. In two-dimensions analogous formulae have been
worked out in \cite{fabbri-navarro-salas-olmo04} and a somewhat
related scheme has been given in \cite{fredenhagen-haag90}.

Let us now apply (\ref{eq:ai+aj}) to the formation process of a
Schwarzschild black hole and restrict the ``out'' region to future
null infinity ($I^+$). The ``in'' region is, as usual, defined by
past null infinity ($I^-$). At $I^+$ we can consider the
normalized radial plane-wave modes $ u^{out}_{wlm}(t,r, \theta,
\phi) = u_{w}(u)r^{-1}Y_{lm}(\theta, \phi)$, where $u_{w}(u)=
\frac{e^{-iwu}}{\sqrt{4\pi w}}$ and $u$ is the null retarded time.
Note that to work with the null hypersurface $I^+$ instead of a
spacelike one requires to replace the two-point function by the
symmetrized one. We shall now evaluate the matrix coefficients
$\langle in|N_{i_1i_2}|in\rangle$ where $i\equiv(w,l,m)$. After
straightforward  manipulations we have
\begin{eqnarray} \label{Nblackhole}&& \langle in|N_{i_1i_2}|in
\rangle = \frac{4}{\hbar}\int_{I^+}du_1d\Omega_1du_2d\Omega_2
Y_{l_1m_1}(\theta_1,\phi_1)\times \nonumber
\\ && Y_{l_2m_2}^*(\theta_2,\phi_2)
u_{w_1}(u_1)
u^*_{w_2}(u_2)\partial_{u_1}\partial_{u_2}[G_{in}(x_1,x_2)\nonumber
\\ &-& G_{out}(x_1,x_2) ] \ \ \ \ ,
\end{eqnarray} where $G_{in}(x_1,x_2)$ and $G_{out}(x_1,x_2)$ are
the two-point functions of the ``in'' and ``out'' states,
respectively. Note that $G_{in}(x_1,x_2)- G_{out}(x_1,x_2)$ is a
smooth function.  The singularity of $G_{in}(x_1,x_2)$ is exactly
cancelled by  the corresponding one of  $G_{out}(x_1,x_2)$. At
$I^+$ these functions can be expanded as
\begin{eqnarray} G_{out}(x_1,x_2) &=&\frac{\hbar}{2}
\int_0^{\infty}dw\sum_{l,m}\frac{e^{-iwu_1}}{\sqrt{4\pi
w}}Y_{lm}(\theta_1,\phi_1) \nonumber
\\ &\times& \frac{e^{iwu_2}}{\sqrt{4\pi w}}Y_{lm}^*(\theta_2,\phi_2) + c.c. \ ,
\end{eqnarray}
and \begin{eqnarray} \label{correlatorin} G_{in}(x_1,x_2)
&=&\frac{\hbar}{2}
\int_0^{\infty}dw\sum_{l,m}\frac{e^{-iwv(u_1)}}{\sqrt{4\pi
w}}Y_{lm}(\theta_1,\phi_1) \nonumber
\\&\times&\frac{e^{iwv(u_2)}}{\sqrt{4\pi
w}}Y_{lm}^*(\theta_2,\phi_2) + c.c. \ , \end{eqnarray} where the
function $v(u)$ in (\ref{correlatorin}) is, as usual, given by \be
\label{relationuv} v \approx constant- \kappa^{-1}e^{-\kappa u} \
. \ee   Note that this expression, relating the inertial times at
$I^+$ and at $I^-$, encodes the effect of the time-dependent
gravitational collapse. Using it assumes that we are in the
late-time regime and also that we are neglecting the backreaction.

Performing first the angular integrations and defining
\begin{eqnarray} \label{shortdistanceout} \tilde{G}_{out}(u_1,u_2)&\equiv
&\hbar\partial_{u_1}\partial_{u_2}\int_0^{\infty}dw\frac{e^{-iw(u_1-u_2)}}{4\pi
w} \nonumber \\&=&-\frac{\hbar}{4\pi}\frac{1}{(u_1-u_2)^2} \ ,
\end{eqnarray} and a similar expression for the ``in'' vacuum
\begin{eqnarray} \label{shortdistancein}
\tilde{G}_{in}(v_1,v_2)=-\frac{\hbar}{4\pi}\frac{1}{(v_1-v_2)^2} \
,
\end{eqnarray}
we easily get
\begin{eqnarray} \label{N12}&&\langle in|N_{i_1i_2}|in \rangle =
 \frac{\hbar^{-1}}{\pi\sqrt{\omega
_1\omega _2}}\int_{I^+}du_1du_2  e^{-i(w_1u_1-w_2u_2)} \nonumber \\
&\times&  \left[
\frac{dv_1}{du_1}\frac{dv_2}{du_2}\tilde{G}_{in}(v_1,v_2)
-\tilde{G}_{out}(u_1,u_2) \right]\delta_{l_1 l_2}\delta_{m_1 m_2}\
\end{eqnarray}
 We can rewrite this expression using
(\ref{relationuv}) and introducing new variables $z^+=u_2+ u_1,
z=u_2-u_1$ so that the integral corresponding to $z^+$ leads to a
delta function in frequencies. The result is
\begin{eqnarray}\label{eq:divergences}
\langle in|N_{i_1 i_2}|in\rangle &=& -\frac{\delta
(w_1-w_2)}{2\pi\sqrt{w_1w_2}} \int_{-\infty}^{+\infty} dz
e^{-i\frac{(w_1+w_2)}{2}z}  \nonumber \\
&\times& \left[ \frac{\kappa^2 e^{-\kappa z}}{(e^{-\kappa z}-1)^2}
-\frac{1}{z^2} \right]\delta_{l_1 l_2}\delta_{m_1 m_2} \ .
\end{eqnarray}
 Finally, performing the integration in $z=u_2-u_1$  we get
the Planckian spectrum (see
(\ref{franja})-(\ref{primitivafranja}))
\begin{eqnarray} \label{integral}\frac{-1}{2\pi w}
\int_{-\infty}^{+\infty} dz e^{-iwz} [ \frac{\kappa^2 e^{-\kappa
z}}{(e^{-\kappa z}-1)^2} -\frac{1}{z^2}] =
\frac{1}{e^{2\pi w \kappa^{-1}}-1}. \end{eqnarray} We note  that,
to obtain this result, we have to assume that quantum field theory
is valid on all scales.

To get the final result we have to take into account the fact that
we have restricted our ``out'' Fock space to the external region
$I^+$. This means that a fraction of an outgoing wave-packet will
be scattered by the potential barrier and only part of it reaches
$I^+$. To incorporate this effect we should multiply the ``out''
modes $u^{out}_{wlm}$ in (\ref{Nblackhole}) by the transmission
coefficients $t_{wl}$ of the Schwarzschild geometry. Therefore we
obtain the complete emission rate per unit frequency $w$ and time
$u$\be \frac{dN}{dwdu}\equiv \frac{1}{2\pi}\langle in
|N_w|in\rangle =
 \frac{1}{2\pi}\frac{\Gamma_{lm}}{e^{2\pi w \kappa^{-1}}-1} \ ,
\ee where $\Gamma_{lm}=|t_{lm}|^2$ are the grey-body coefficients.
When the black hole is rotating the result is similar to
(\ref{integral}) with the replacement of $w$ by $w-m\Omega_H$. The
analysis can also be extended to account for correlations between
number operators with different frequencies. They can be expressed
as \cite{olmo05} $\langle in|N_{i_1} N_{i_2}|in\rangle - \langle
in|N_{i_1}|in\rangle \langle in|N_{i_2}|in\rangle =  |\langle
in|N_{i_1 i_2}|in\rangle |^2 + |\langle in|C_{i_1
i_2}|in\rangle|^2$, where $C_{i_1 i_2}$ is the operator
\begin{eqnarray}
C_{i_1i_2}&=& \int_\Sigma d\Sigma_1 ^\mu d\Sigma_2 ^\nu
[u^{out*}_{i_1}(x_1){\buildrel\leftrightarrow\over{\partial}}_\mu
][u^{out*}_{i_2}(x_2){\buildrel\leftrightarrow\over{\partial}}_\nu
]\nonumber \\ &\times& (\phi (x_1)\phi (x_2)- \langle out|\phi
(x_1)\phi (x_2)|out\rangle) \ . \end{eqnarray} Explicit evaluation
gives
\begin{eqnarray} \label{C12}&&\langle in|C_{i_1i_2}|in \rangle =
-\frac{\delta (w_1+w_2)}{2\pi\sqrt{w_1w_2}}
\int_{-\infty}^{+\infty} dz
e^{-i\frac{(w_2-w_1)}{2}z}  \nonumber \\
&\times& \left[ \frac{\kappa^2 e^{-\kappa z}}{(e^{-\kappa z}-1)^2}
-\frac{1}{z^2} \right]\delta_{l_1 l_2}\delta_{m_1 m_2} \ .
\end{eqnarray}
We note  that the  behavior of the two-point functions
(\ref{shortdistanceout}) and (\ref{shortdistancein}) (both
$\frac{dv_1}{du_1}\frac{dv_2}{du_2}\tilde{G}_{in}(v_1,v_2)$ and
 $\tilde{G}_{out}(u_1,u_2)$ can be expressed in terms of $z=u_2-u_1$) is
fundamental for the vanishing of both quantities $\langle
in|C_{i_1 i_2}|in\rangle $ and $\langle in|N_{i_1 i_2}|in\rangle$
(the latter with $i_1 \neq i_2$).

 The expression (\ref{integral}) is very useful since it offers
an explicit way to evaluate the ``weight'' of distances
$|u_2-u_1|$ to the Planckian spectrum. To be more explicit we
shall now compute the contribution of distances $z \in [-\epsilon,
\epsilon]$ to the full integral. This contribution
\begin{eqnarray}\label{franja} I(w,\kappa,\epsilon)&=& \frac{-1}{2\pi w}
\int_{-\epsilon}^{+\epsilon} dz e^{-iwz}
[ \frac{\kappa^2 e^{-\kappa z}}{(e^{-\kappa z}-1)^2}
-\frac{1}{z^2}] \ \  \end{eqnarray} can be evaluated analytically
\begin{eqnarray} \label{primitivafranja}& & I(w,\kappa,\epsilon)=1 -\frac{1}{2 \pi w} \{
iw [-\frac{2\kappa}{w}\sin{w\epsilon} - i\pi-2i Si
(w\epsilon)+\nonumber\\ && i \frac{\kappa e^{-i w \epsilon}}{w}
\{_2F_1[1,-\frac{i \omega}{\kappa},1-\frac{i
\omega}{\kappa},e^{\kappa \epsilon}]\}- \nonumber \\ &&i
\frac{\kappa e^{iw\epsilon}}{w} \{_2F_1[1,-\frac{i
w}{\kappa},1-\frac{i w}{\kappa},e^{-\kappa \epsilon}]\}]+
\frac{e^{-i \epsilon w}}{\epsilon (e^{\epsilon \kappa}-1)} \times
\nonumber \\ &&  [ -1+e^{\epsilon(\kappa+2i w)}+e^{\epsilon
\kappa} (1-\epsilon \kappa)-e^{2 i \epsilon w}(1+\epsilon\kappa)]
\}  \ , \end{eqnarray} and in the limit $\epsilon \to +\infty$
  we nicely recover  the Planckian spectrum
$(e^{2\pi w \kappa^{-1}}-1)^{-1}$. We note that the above
expression holds equally for an arbitrary number $4+n$ of
dimensions. Moreover, a simple calculation shows that the absence
of correlations $\langle in|N_{i_1} N_{i_2}|in\rangle - \langle
in|N_{i_1}|in\rangle \langle in|N_{i_2}|in\rangle =0$ in the
emitted radiation is preserved even if short distances are
excluded in the evaluation of $\langle in|C_{i_1 i_2}|in\rangle $
and $\langle in|N_{i_1 i_2}|in\rangle$.

For black holes produced by gravitational stellar collapse the
contribution of $I(w,\kappa,\epsilon)$ is, when $\epsilon$ is
taken as the  Planck length $l_P= 1.6\times 10^{-33}cm $,
negligible (of order $\kappa \epsilon$ for $w_{typical} \sim
\kappa/2\pi \equiv T_H$). In fact,  for a black hole of three
solar masses we need high frequencies $w/w_{typical}\approx 96$ to
find that the contribution of Planck distances $I(w, \kappa, l_P)$
is of order of the total spectrum itself.  Moreover, the relative
contribution to the Planckian distribution is, for
$w=w_{typical}$, of order $10^{-38}\%$. For primordial black holes
$M\approx 10^{15}g$ we find $w/w_{typical}\approx 52$ and the
relative contribution to the spectrum is now $10^{-19}\%$. This is
why Hawking thermal radiation is very robust, as it has been
confirmed in analysis based on acoustic black holes (for recent
reviews, see \cite{reviews}). The condition on $|u_1-u_2|$, which
accounts for very short wavelength, is analogous to the
modification of the dispersion relations in the fluid frame. The
deviations from the Planckian spectrum are also found, in acoustic
black holes, of order $\kappa k_0$  ($k_0$ is the wave vector
characterizing the fluid atomic scale) for $w\sim w_{typical}$.

 When the product $\epsilon\kappa$ is of unit order the
contribution of short distances to the Planckian spectrum is not
negligible. The integral $I(w,\kappa, \epsilon)$ gives values
similar to $(e^{2\pi w \kappa^{-1}}-1)^{-1}$ when $w/w_{typical}$
is not very high. This happens in TeV gravity scenarios. Assuming
a drastic change of the strength of gravity at short distances due
to $n$ extra dimensions (a Planck mass $M_{TeV}$ of 1 TeV) and for
a $(4+n)$-dimensional Schwarzschild black hole of mass $M$ ($M\sim
5-10$ TeV) with \cite{myers-perry} $\kappa = \frac{(n+1)}{2r_H}$,
where the horizon radius is given by \be r_H=
\frac{2}{M_{TeV}}\left (\frac{M}{M_{TeV}}\right
)^{\frac{1}{n+1}}\left
(\frac{\pi^{(n-3)/2}\Gamma((n+3)/2)}{n+2}\right )^{\frac{1}{n+1}}
\nonumber , \ee
\begin{figure}[htbp]
\begin{center}
\includegraphics[angle=0,width=1.8in,clip]{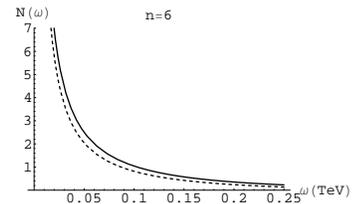}
\caption{Plot comparing the Planckian distribution (solid line)
$N(w,\kappa)=(e^{2\pi w/ \kappa}-1)^{-1}$ with the one obtained by
suppressing the contributions coming from distances shorter than
$\epsilon=l_{TeV}$ (dotted line). We have taken $M=10$ TeV.
}\label{Fig1}
\end{center}
\end{figure}
we obtain:
 $w/w_{typical} \approx 3.3$ ($n=2$), $w/w_{typical} \approx 3.1$ ($n=4$) and $w/w_{typical} \approx 3.0$
 ($n=6$), for a black hole mass $M=5$ TeV; $w/w_{typical} \approx 3.6$ ($n=2$), $w/w_{typical} \approx 3.3$ ($n=4$) and
 $w/w_{typical} \approx 3.1$
 ($n=6$), for $M=10$ TeV.
 $w_{typical}$ varies in the interval $\sim 100-165$ GeV, depending on $n$ and $M$. The contribution of distances shorter
than the new Planck length $l_{TeV}\sim 10^{-17} cm$ to the
spectrum reaches now significative values: $21\%$ ($n=2$), $25\%$
($n=4$) and $28\%$ ($n=6$) for $M=5$ TeV, and $17\%$ ($n=2$),
$22\%$ ($n=4$) and $26\%$ ($n=6$) for $M=10$ TeV (see Fig. 1). In
addition, the relative contribution to the luminosity, originated
in the distance range $|u_2 - u_1|< l_{TeV}$, increases these
numbers since grey-body factors $\Gamma_{l}(w)$ grow up with
frequency. Since in the ultrashort distance regime there may exist
some unknown physics, not described by relativistic quantum field
theory, it can give  some signature in the evaporation, even
before reaching the Planck-scale phase. In other words,
significant deviations from the Planckian spectrum can potentially
emerge in the ``Schwarzschild phase''  of the evaporation, where
most of the energy is expected to be radiated away
\cite{giddings-thomas02}.

Finally we wish to stress  that Eq. (\ref{N12}) can be rewritten
as an integral along $I^-$ (with respect to $dv_1dv_2$).
Constraining distances also at $I^-$ in the ``naive'' way:
$(v_2-v_1)^2 \sim \kappa^{-2} (e^{-\kappa u_2}-e^{-\kappa u_2})^2
< \epsilon$ is problematic. To see this let us consider Minkowski
space and the transformation $ v=e^{-\xi}u$, which can be regarded
as a radial boost
 with rapidity $\xi$. Absence of particle production under this
 boost requires that, at $I^-$, we should
impose $(v_2-v_1)^2<\epsilon^2 e^{-2\xi}$ (if
$(u_2-u_1)^2<\epsilon^2$) or $(u_2-u_1)^2<\epsilon^2 e^{2\xi}$ (if
$(v_2-v_1)^2<\epsilon^2$) . Therefore, under a general
transformation $v=v(u)$ (as the one $v \sim \kappa^{-1} e^{-\kappa
u}$ appearing in black hole formation) we should generalize the
above relations and the easiest way is $(v_2-v_1)^2<\epsilon^2
\frac{dv_1}{du_1}\frac{dv_2}{du_2}$ (if $(u_2-u_1)^2<\epsilon^2$)
or $(u_2-u_1)^2<\epsilon^2 \frac{du_1}{dv_1}\frac{du_2}{dv_2}$ (if
$(v_2-v_1)^2<\epsilon^2$). In the former situation (naturally
preferred since physical measurements are performed at $I^+$) the
results are equivalent to those obtained previously and parallel
to those obtained in sonic black holes. The second possibility is
more exotic since it predicts a drastic change in the particle
production rate \cite{preparation}.  The radiation is
approximately thermal after the formation of the black hole, but
for a short period. Moreover, the correlations cease to be zero
and increase with time. This possibility cannot be excluded
completely (see also \cite{unruh-schutzhold}).

We thank A. Fabbri and L. Parker for useful comments and
suggestions. I.A. thanks MEC for a FPU fellowship. This work has
been  partially supported by grants FIS2005-05736-C03-03 and EU
network MRTN-CT-2004-005104. G.J.O. has been supported by NSF
grants PHY-0071044 and PHY-0503366.


\begin{thebibliography}{99}
\bibitem{hawk1} S. W. Hawking,  {\it Comm. Math. Phys.} {\bf 43}
199 (1975)
\bibitem{hooft} G.'t Hooft, {\it Nucl. Phys.} {\bf B335}, 138 (1990)
\bibitem{jacobson9193} T. Jacobson, {\it Phys. Rev.} D {\bf 44}
1731 (1991); {\it Phys. Rev} D {\bf 48} 728 (1993)
\bibitem{unruh95} W.G. Unruh, {\it Phys. Rev.} D {\bf 51} 2827
(1995)
\bibitem{bmps-cj} R. Brout, S. Massar, R. Parentani and P. Spindel,
{\it Phys. Rev.} D {\bf 52} 4559 (1995); S. Corley and T.
Jacobson, {\it Phys. Rev.} D {\bf 54} 1568 (1996); {\it Phys.
Rev.} D {\bf 59} 124011 (1999); S. Corley, {\it Phys. Rev.} D {\bf
57} 6280 (1998)


\bibitem{giddings-thomas02} S.B. Giddings and S. Thomas, {\it Phys.
Rev.} D {\bf 65}, 056010 (2002). S.B. Giddings, {\it
Gen.Rel.Grav.} {\bf 34}, 1775 (2002). B.J. Carr and S.B. Giddings,
{\it Sci.Am.} {\bf 292} N5:30-37 (2005)
\bibitem{dimopoulos-landsberg02} S. Dimopoulos and G. Landsberg,
{\it Phys. Rev. Lett.} {\bf 87}, 161602 (2001).

\bibitem{bhaccelerators} S. Dimopoulos and R. Emparan, {\it Phys. Lett.} B {\bf 526},
393 (2002). G. Landsberg, {\it Phys. Rev. Lett.} {\bf 88} 181801
(2002). C.M. Harris and P. Kanti, {\it JHEP} 0310:014 (2003). J.L.
Hewett, B. Lillie and T. Rizzo, {\it Phys. Rev. Lett.} {\bf 95}
261603 (2005)

\bibitem{extradimensions} N. Arkani-Hamed, S.
Dimopoulos and G.R. Dvali, {\it Phys. Lett.} B {\bf 429}, 263
(1998). I. Antoniadis, N. Arkani-Hamed, S. Dimopoulos and G.R.
Dvali, {\it Phys. Lett.} B {\bf 436}, 257 (1998). L. Randall and
R. Sundrum, {\it Phys. Rev. Lett.} {\bf 83}, 3370 (1999)
\bibitem{banks-fischler99} T. Banks and W. Fischler,
hep-th/9906038
\bibitem{emparan-horowitz-myers} R. Emparan, G.T. Horowitz and
R.C. Myers, {\it Phys. Rev. Lett.} {\bf 85} 499 (2000)
\bibitem{parker75}L. Parker  {\it Phys. Rev.} D {\bf 12}, 1519 (1975)
\bibitem{wald75} R. M. Wald {\it Commun. Math. Phys.} {\bf 45}, 9
(1975)

\bibitem{waldbook} R. M. Wald  {\it Quantum field theory in curved spacetime and
black hole thermodynamics}, CUP, Chicago (1994)

\bibitem{icp05} A. Fabbri  and J. Navarro-Salas  {\it Modeling black hole evaporation},
ICP-World Scientific, London (2005)

\bibitem{fabbri-navarro-salas-olmo04} A. Fabbri, J. Navarro-Salas
and G.J. Olmo, {\it Phys. Rev.} D {\bf 70} 064022 (2004)

\bibitem{fredenhagen-haag90} K. Fredenhagen and R. Haag, {\it
Commun. Math. Phys.} {\bf 127} 273 (1990)
\bibitem{olmo05} G.J. Olmo,  Ph.D Thesis, University of Valencia
 (2005)
 \bibitem{reviews} C. Barcel\'{o}, S. Liberati and M. Visser, {\it
 Living
Rev. Rel.} {\bf 8}:12 (2005); R. Balbinot, A. Fabbri, S. Fagnocchi
and R. Parentani, Riv. Nuovo Cimento {\bf 28}, 1 (2005)
gr-qc/0601079
\bibitem{myers-perry} R.C.Myers and M.J.Perry,{\it Ann.Phys.}(N.Y.){\bf 172},304(1986)
\bibitem{preparation} Work in preparation
\bibitem{unruh-schutzhold} W.G. Unruh and R. Schutzhold, {\it
Phys. Rev.} D {\bf 71}, 024028 (2005)



\end{thebibliography}
\end{document}